\documentclass[10pt,letterpaper]{article}
\usepackage[top=0.85in,left=2.75in,footskip=0.75in]{geometry}
\usepackage{float}

\usepackage{amssymb}

\usepackage{amsmath}
\DeclareMathOperator*{\argmax}{arg\,max}

\DeclareMathOperator{\asinh}{asinh}

\usepackage{changepage}

\usepackage[utf8x]{inputenc}

\usepackage{textcomp,marvosym}

\usepackage{cite}

\usepackage[right]{lineno}
\usepackage{microtype}
\usepackage{enumitem}

\DisableLigatures[f]{encoding = *, family = * }

\usepackage[x11names, table]{xcolor}

\usepackage{array}

\usepackage{acronym}
\usepackage{etoolbox}

\usepackage{tikz}
\usetikzlibrary{fit,positioning,arrows,automata,calc}
\tikzset{
  main/.style={rectangle, minimum size = 10mm, thick, draw =black!80, node distance = 15mm},
  connect/.style={-latex, thick},
  box/.style={rectangle, draw=black!100}
}

\tikzstyle{hidden}=[shape=rectangle,draw=black!80,fill=white!80,minimum size = 10mm, thick,node distance = 15mm]
\tikzstyle{observation}=[shape=circle,draw=black!80,fill=black!10,minimum size = 10mm, thick,node distance = 15mm]

\tikzstyle{state}=[shape=circle,draw=black!80,fill=white!80,minimum size = 10mm, thick,node distance = 10mm]

\usepackage{newfloat}
\usepackage{mdframed}
\usepackage{subcaption}

\usepackage[retain-unity-mantissa=false,binary-units=true]{siunitx}

\DeclareSIUnit\basepair{bp}

\usepackage{multicol}
\usepackage[mathcal]{eucal}
\usepackage{booktabs}
\usepackage{tabularx}
\usepackage{adjustbox}

\usepackage[]{todonotes}
\reversemarginpar{}
\usepackage[normalem]{ulem}

\newcolumntype{+}{!{\vrule width 2pt}}

\newlength{\savedwidth}

\raggedright{}
\usepackage[aboveskip=1pt,labelfont=bf,labelsep=period,justification=raggedright,singlelinecheck=off]{caption}

\makeatletter
\renewcommand{\@biblabel}[1]{\quad#1.}
\makeatother

\usepackage{lastpage,fancyhdr,graphicx}
\usepackage{epstopdf}
\fancyheadoffset[L]{2.25in}
\fancyfootoffset[L]{2.25in}
\usepackage{longtable}
\usepackage{caption}

\usepackage{nameref}
\usepackage[hidelinks]{hyperref}

\DeclareFloatingEnvironment[name=Box,placement=htbp]{textboxfloat}
\newmdenv[backgroundcolor=black!20,linewidth=0pt,innerleftmargin=1em,innerrightmargin=1em,innertopmargin=1em,innerbottommargin=1em]{textboxframe}

\newcommand*{\orcidauthor}[2]{\mbox{#1}~(\href{https://orcid.org/#2}{#2})}

\begin{document}
\vspace*{0.2in}

% A review should aim for 3000-6000 words and two or three figures or other display items. 2020-09-02 Overleaf word count: 5976. % http://journals.plos.org/ploscompbiol/s/other-article-types

\begin{flushleft}
  {\Large
    \textbf\newline{Segmentation and genome annotation algorithms} % Please use "sentence case" for title and headings (capitalize only the first word in a title (or heading), the first word in a subtitle (or subheading), and any proper nouns).
  }
  \newline
  \\
  \orcidauthor{Maxwell W.~Libbrecht}{0000-0003-2502-0262}\textsuperscript{1{\Yinyang}*}, % chktex 8
  \orcidauthor{Rachel C.W.~Chan}{0000-0003-1009-6379}\textsuperscript{2,3\Yinyang}, % chktex 8
  \orcidauthor{Michael M.~Hoffman}{0000-0002-4517-1562}\textsuperscript{2,3,4,5*} % chktex 8
  \\
  \bigskip
  \textbf{1} School of Computing Science, Simon Fraser University, Burnaby, BC, Canada
  \\
  \textbf{2} Department of Computer Science, University of Toronto, Toronto, ON, Canada
  \\
  \textbf{3} Princess Margaret Cancer Centre, Toronto, ON, Canada
  \\
  \textbf{4} Department of Medical Biophysics, University of Toronto, Toronto, ON, Canada
  \\
  \textbf{5} Vector Institute for Artificial Intelligence, Toronto, ON, Canada
  \\
  \bigskip

  % Insert additional author notes using the symbols described below. Insert symbol callouts after author names as necessary.
  %
  % Remove or comment out the author notes below if they aren't used.
  %
  % Primary Equal Contribution Note
  {\Yinyang} These authors contributed equally to this work.

  % Additional Equal Contribution Note
  % Also use this double-dagger symbol for special authorship notes, such as senior authorship.
  % \ddag These authors also contributed equally to this work.

  % \textcurrency Current Address: Dept/Program/Center, Institution Name, City, State, Country % change symbol to "\textcurrency a" if more than one current address note
  % \textcurrency b Insert second current address
  % \textcurrency c Insert third current address

  % \dag Deceased

  % Group/Consortium Author Note
  % \textpilcrow Membership list can be found in the Acknowledgments section.

  % Use the asterisk to denote corresponding authorship and provide email address in note below.
  * maxwell\_libbrecht@sfu.ca, michael.hoffman@utoronto.ca

\end{flushleft}

\section*{Abstract}
\Ac{SAGA} algorithms are widely used to understand genome activity and gene regulation.
These algorithms take as input epigenomic datasets, such as \ac{ChIP-seq} measurements of histone modifications or transcription factor binding.
They partition the genome and assign a label to each segment such that positions with the same label exhibit similar patterns of input data.
\Ac{SAGA} algorithms discover categories of activity such as promoters, enhancers, or parts of genes without prior knowledge of known genomic elements.
In this sense, they generally act in an unsupervised fashion like clustering algorithms, but with the additional simultaneous function of segmenting the genome.
Here, we review the common methodological framework that underlies these methods, review variants of and improvements upon this basic framework, catalogue existing large-scale reference annotations, and discuss the outlook for future work.

% Please keep the Author Summary between 150 and 200 words
% Use first person. PLOS ONE authors please skip this step.
% Author Summary not valid for PLOS ONE submissions.
% \section*{Author summary}
% Lorem ipsum dolor sit amet, consectetur adipiscing elit.

%\linenumbers{}

\acresetall{}

\section*{Background and motivation}

High-throughput sequencing technology has enabled numerous techniques for genome-scale measurement of chemical and physical properties of chromatin and associated molecules in individual cell types.
Using sequencing assays, the \ac{ENCODE} Project, the Roadmap Epigenomics Project, and myriad individual researchers have generated thousands of such datasets.
These datasets quantify various facets of gene regulation such as genome-wide transcription-factor binding, histone modifications, open chromatin, and RNA transcription.
Each dataset measures a particular activity at billions of positions, and the collection of datasets does so in hundreds of samples across a variety of species and tissues.
Transforming these quantifications of diverse properties into a holistic understanding of each part of the genome requires effective means for summarization.
\Ac{SAGA} algorithms~(\autoref{box:saga}) have emerged as the predominant way to summarize activity at each position of the genome, distilling complex data into an interpretable précis of genomic activity.

\begin{textboxfloat}
  \begin{textboxframe}

    \caption{Terminology}\label{box:saga}

    \textbf{SAGA.}
    We define a \acf{SAGA} algorithm as a procedure that:
    \begin{enumerate}[noitemsep]
    \item assigns to each position of a whole genome a label (``genome annotation''),
    \item from a set of multiple (\num{\ge 3}) classes, % chktex 1
    \item by
      \begin{enumerate}[nosep]
      \item integrating multiple independent observations at each position, and
      \item modeling dependence between adjacent positions (``segmentation'').
      \end{enumerate}
    \end{enumerate}
    Previously, researchers have used several other terms to describe this task, including ``segmentation''~\cite{day2007unsupervised}, ``chromatin state annotation''~\cite{ernst2012chromhmm} and ``semi-automated genome annotation''~\cite{libbrecht2015joint}.

    \setlength{\parskip}{\baselineskip}%
    \noindent {\hrulefill}
    % to separate SAGA from the other alphabetically listed terms

    \textbf{Assay.}
    An experiment that produces a measurement at each genomic position, such \ac{ChIP-seq} or \ac{ATAC-seq}.

    \textbf{Label.}
    One of a finite set of classes assigned to each genomic segment that shares similar activity.
    Other terms include ``state'' or ``chromatin state''.

    \textbf{Sample.}
    A population of cells on which one can perform an assay, such as a primary tissue sample or a cell line.
    Other terms include ``cell type'', ``epigenome'', or ``biosample''.

    % (1)~Epigenomics datasets. Other names: functional genomics datasets, sequencing-based genomics datasets, genomics datasets.
    % (2)~Dataset. Other names: track.
    % (3)~\ac{SAGA} algorithm. Other names: chromatin state annotation, unsupervised genome annotation, genome segmentation.
    % (4)~Label. Other names: state, chromatin state, chromatin color.
    % (5)~Interpretation term. Other names: mnemonic.
    % (6)~Sample. Other names: cell type, epigenome, cell line, biosample.

  \end{textboxframe}
\end{textboxfloat}

\begin{figure}[htbp]
  \centering
  \includegraphics[width=\textwidth]{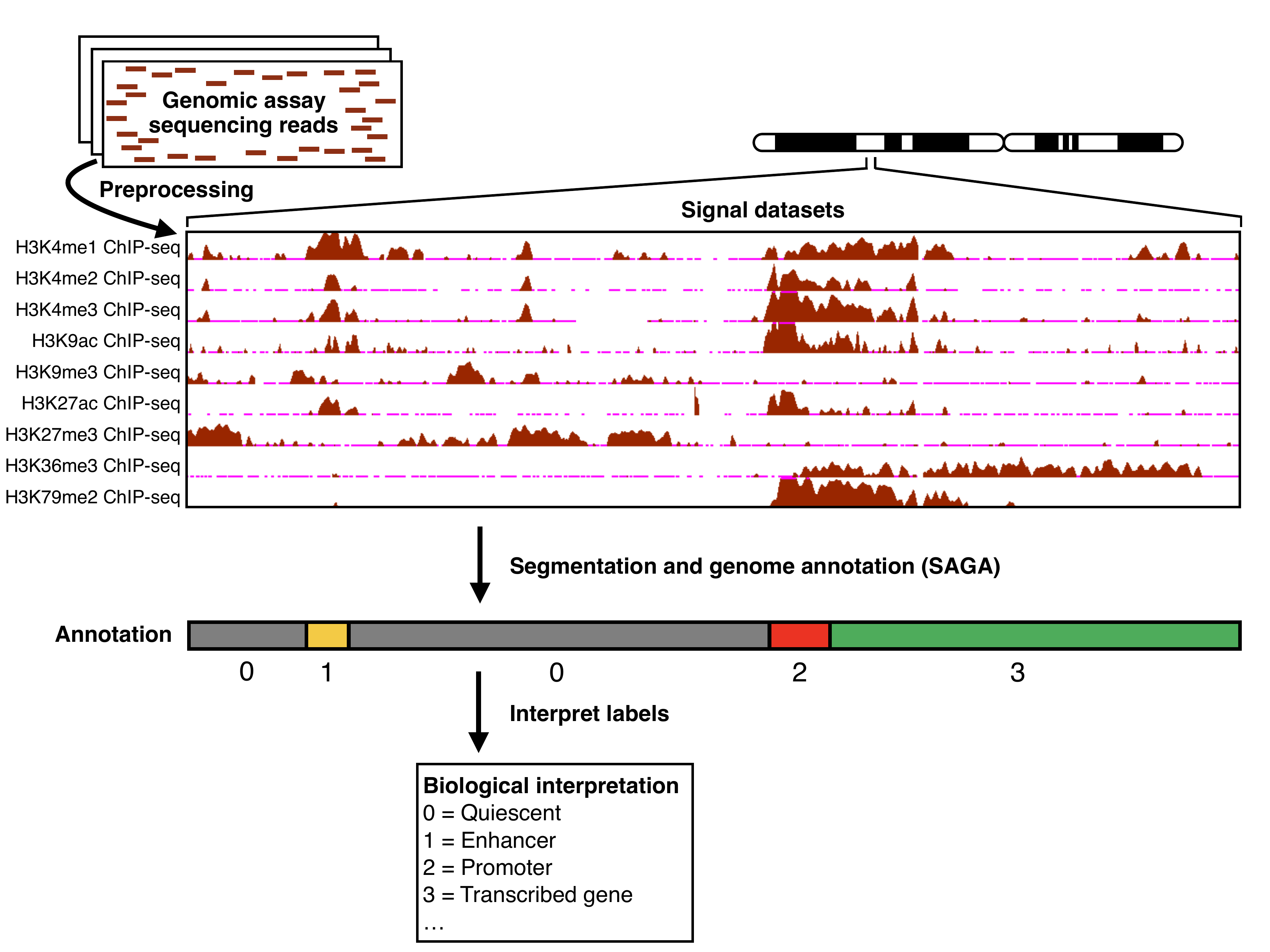}
  \caption{\textbf{Overview of \acf{SAGA}.}
    First, preprocessing transforms genomic assay sequencing reads into signal datasets.
    Second, with signal datasets as input, a \ac{SAGA} algorithm partitions the genome and assigns an integer label to each segment, yielding an annotation.
    Third, a researcher interprets the labels, assigning a biological interpretation to each.}\label{fig:saga}
% Caption needs more detail. Should explain the origin of the ENCODE color scheme.}
\end{figure}

\Ac{SAGA} algorithms take as input a collection of genomic datasets, such as \ac{ChIP-seq} measurements of histone modifications or of transcription factor binding~(\autoref{fig:saga}).
The \ac{SAGA} task is to use the input datasets to partition the genome into segments and assign a label to each segment.
\Ac{SAGA} algorithms perform this task in a way that leads to positions with the same label having similar patterns in the input data.

% https://tex.stackexchange.com/questions/31441/vertical-line-only-in-one-part-of-a-table
\newcommand{\junction}{\hfil\kern\arraycolsep\vline\hspace{-0.26em}\textbullet\kern-\arraycolsep\hfilneg}

\begin{table}[htbp]
  \begin{tabularx}{\textwidth}{r X r} % chktex 44
    \hline % chktex 44
    \textbf{Year} \phantom{\junction} & \textbf{Name or description} & \textbf{References} \\
    \hline % chktex 44
    2007 {\junction} & HMMSeg & \cite{day2007unsupervised} \\ % chktex 2
    2010 {\junction} & Chromatin colors & \cite{filion2010systematic} \\ % chktex 2
    2010 {\junction} & Chromatin states model & \cite{ernst2010discovery} \\ % chktex 2
    2012 {\junction} & ChromHMM & \cite{ernst2012chromhmm, hoffman2012integrative, kundaje2015integrative, ernst2017chromatin} \\ % chktex 2
    2012 {\junction} & Segway & \cite{hoffman2012unsupervised, hoffman2012integrative, chan2018segway, libbrecht2019unified} \\ % chktex 2
    2013 {\junction} & TreeHMM & \cite{biesinger2013discovering} \\ % chktex 2
    2015 {\junction} & Spectacle & \cite{song2015spectacle} \\ % chktex 2
    % 2015 & Roadmap Epigenomics (with ChromHMM) & \cite{kundaje2015integrative} \\ % chktex 2
    2015 {\junction} & hiHMM & \cite{sohn2015hihmm} \\ % chktex 2
    2015 {\junction} & Ensembl Regulatory Build (with Segway, ChromHMM) & \cite{zerbino2015ensembl} \\ % chktex 2
    2015 {\junction} & EpiCSeg & \cite{mammana2015chromatin} \\ % chktex 2
    2015 {\junction} & Segway+GBR & \cite{libbrecht2015joint, libbrecht2015entropic} \\ % chktex 2
    2016 {\junction} & IDEAS & \cite{zhang2016jointly, zhang2017accurate} \\ % chktex 2
    2017 {\junction} & GenoSTAN & \cite{zacher2017accurate} \\ % chktex 2
    2017 {\junction} & diHMM & \cite{marco2017multi} \\ % chktex 2
    2018 {\junction} & iSeg & \cite{girimurugan2018iseg} \\ % chktex 2
    2018 {\junction} & StatePaintR & \cite{coetzee2020statehub} \\ % chktex 2
    2019 {\junction} & RT States & \cite{poulet2019rt} \\ % chktex 2
    2019 {\junction} & ConsHMM & \cite{arneson2019systematic} \\ % chktex 2
    2020 {\junction} & modHMM & \cite{benner2020modhmm} \\ % chktex 2
    2020 {\junction} & SPIN & \cite{wang2020spin} \\ % chktex 2
    2020 {\junction} & SegRNA & \cite{mendez2020unsupervised} \\ % chktex 2
    \hline % chktex 44
  \end{tabularx}
  \caption{\textbf{Timeline of selected \ac{SAGA} methods.}}\label{table:SAGAtimeline}
\end{table}

The first \ac{SAGA} methods were developed in the 2000s, but have increased in usage recently, thanks to the wide availability of genomic datasets~(\autoref{table:SAGAtimeline}).
Large-scale genomic profiling projects such as \ac{ENCODE}~\cite{encode2012integrated} and Roadmap Epigenomics~\cite{kundaje2015integrative} produced \ac{SAGA} annotations as a primary output.
Researchers have developed a large variety of \ac{SAGA} strategies with the goal of improving upon the basic \ac{SAGA} framework.

In this review, we summarize the main strategies used by most \ac{SAGA} methods.
Then, we discuss differences between methods, the challenges they face, and the outlook for future work.

\section*{Input data}\label{section:inputdata}

\subsection*{Experimental assays used for input data}

\Ac{SAGA} methods typically use as input a number of different experimental datasets, each describing some local property of the genome~\cite{zitnik2018review}.
Such properties might include chromatin accessibility or presence of some DNA-binding protein.
Although input data initially came from microarray methods such as tiling arrays~\cite{encode2007identification}, they now usually come from sequence census assays~\cite{wold2007sequence}.

A standard collection of input datasets might measure histone modifications or DNA-binding proteins (using assays like \ac{ChIP-seq}~\cite{barski2007high} or \ac{CUTnRUN}~\cite{skene2017efficient}), and chromatin accessibility (using assays like \ac{DNase-seq}~\cite{boyle2008high, hesselberth2009global} or \ac{ATAC-seq}~\cite{buenrostro2013transposition}).
Supplying \iac{SAGA} algorithm with datasets that measure chromatin activity yields an output annotation that captures the regulatory state of chromatin.
Creating these chromatin activity annotations has served as the predominant use of \ac{SAGA} methods thus far.

Less frequently, researchers have gone beyond measurements of chromatin and DNA-binding proteins and have used \ac{SAGA} methods for other kinds of data.
The output annotation summarizes the input datasets, so the choice of input greatly influences the annotation's content and its subsequent interpretation.
\ac{SAGA} methods can work for any sort of dense linear signal along the genome.
Individual studies have applied it DNA replication timing data~\cite{libbrecht2015joint, poulet2019rt, wang2020spin}, interspecies comparative genomics data~\cite{arneson2019systematic}, and RNA-seq data~\cite{mendez2020unsupervised}.
Other studies have even found ways to incorporate non-linear chromatin 3D genome organization data into the \ac{SAGA} framework~\cite{libbrecht2015joint, wang2020spin}.

\subsection*{Signal representation of genomic assays}

Most genomic assay data so far has come from bulk samples of cells.
These data depict a noisy mixture of sampling an assayed property from the many cells within the population.
These cells may represent subpopulations of slightly different types, or within different cell cycle stages.
Thus, each subpopulation might have different characteristics in the assayed properties.
In the mixture of cell subpopulations, only frequently sampled properties will rise above background noise.
By comparison, less frequently sampled properties seen in a minority of cells, may remain indistinguishable from background noise.

Often, the property examined by an epigenomic assay is exhibited or not exhibited by some position of a single chromosome in a single cell, with no gradations between the extremes.
For example, at some nucleotide of one chromosome in a single cell, an interrogated histone modification is either present or it is not.
A single diploid cell has two copies of the chromosome.
Thus, at that position, each eudiploid cell can have only 0, 1, or 2~instances of the histone modification.

Summing or averaging discrete counts over a population of cells leads to a representation of the assay data called ``signal''.
Signal appears as a continuous-scale measurement.
Signal arises, however, only from the aggregation of position-specific properties, which in each cell may have only a small number of potential ordinal value at the moment of observation.

Unlike epigenomic assays, transcriptomic assays can measure any number of transcript copies of one position per cell.
Despite similar data representations, one must avoid the temptation to interpret epigenomic signal intensity as one might interpret transcriptomic signal intensity.
For a transcriptomic assay, greater signal intensity for a transcriptomic assay might reflect a greater ``level'' of some transcriptional property within each cell.
For an epigenomic assay, greater signal intensity indicates primarily that a higher number of cells within a sample have the property of interest.

In both the epigenomic and transcriptomic cases, it remains difficult or impossible to untangle the contribution to higher signal intensity that arises from frequency of molecular activity within each cell of a subpopulation from that from the composition of subpopulations within a whole bulk population.
Improvements in single-cell assays, however, may enable \ac{SAGA} algorithms on data from single cells in the near future (see ``Outlook for future work'').

\subsection*{Preprocessing of input data}

\Ac{SAGA} methods generally use a signal representation of the input data.
This signal representation originates from raw experimental data, such as sequencing reads, by way of a preprocessing procedure.
For simplicity, we describe the steps of preprocessing as if a human analyst conducted them all individually, although some \ac{SAGA} software packages might perform some steps without manual intervention:

\begin{itemize}[label={},align=left,leftmargin=1.5em,labelwidth=0pt,labelsep=0pt]
\item Required preprocessing for all \ac{SAGA} methods:

  \begin{enumerate}[series=majorpreproc,align=left,leftmargin=1.5em,labelwidth=1.5em]
  \item The analyst transforms the experimental data into raw numeric signal data.
    \begin{itemize}[nosep,label=\textbullet,align=left,leftmargin=1.5em,labelwidth=1.5em]
    \item For sequencing data, the analyst:
      \begin{enumerate}[nosep,label=\arabic*.,align=left,leftmargin=1.5em,labelwidth=1.5em]
      \item Aligns each sequencing read to the reference genome,
      \item May choose to extend each read to an estimated length of the DNA fragment it begins, and
      \item Computes the number of reads per base or extended reads per base for each genomic position~\cite{hoffman2012unsupervised, chan2018segway}.
      \end{enumerate}
    \item For microarray data, the analyst:
      \begin{enumerate}[nosep,label=\arabic*.,align=left,leftmargin=1.5em,labelwidth=1.5em]
      \item Acquires microarray signal intensity for the experimental sample and for a control sample, and
      \item Computes the ratio of experimental intensity to control intensity.
      \end{enumerate}
    \end{itemize}

  \item The analyst chooses units to represent the strength of activity at each position and may perform further transformation of the raw numeric signal data into these units.
    \begin{itemize}[nosep,label=\textbullet,align=left,leftmargin=1.5em,labelwidth=1.5em]
    \item For sequencing data, the analyst commonly uses one of:
      \begin{itemize}[nosep,label=\textbullet,align=left,leftmargin=1.5em,labelwidth=1.5em]
      \item Read count (no transformation),
      \item Fold enrichment of observed data relative to a control~\cite{hoffman2012integrative}, or
      \item $-\log_{10}$ Poisson p-values indicating the likelihood of statistically significant peaks relative to control~\cite{zhang2017accurate}.
      \end{itemize}
      The latter two units can mitigate experimental artifacts because they compare to a control experiment such as a ChIP input control.
    \item For microarray data, the analyst commonly performs $\log_2$ transformation of the intensity ratios~\cite{schuettengruber2009functional, filion2010systematic, kharchenko2011comprehensive}.
    \end{itemize}
  \end{enumerate}

\item Optional preprocessing or preprocessing required only for specific \ac{SAGA} methods:
  \begin{enumerate}[resume=majorpreproc,align=left,leftmargin=1.5em,labelwidth=1.5em]
  \item The analyst may normalize data to harmonize signal across cell types~\cite{xiang2020s3norm}.
    Normalization proves especially important when annotating multiple cell types (see ``Annotating multiple cell types'').

  \item
  To prevent large outlier signal values from dominating the results, the analyst may transform signals using one of three variance-stabilizing transformations of each signal value~$x$:
    \begin{itemize}[nosep,label=\textbullet,align=left,leftmargin=1.5em,labelwidth=1.5em]
    \item $\asinh x$~\cite{hoffman2012unsupervised},
    \item $\log_2 (x+\text{pseudocount})$~\cite{zhang2017accurate}, or
    \item an empirical variance-stabilizing transformation~\cite{bayat2020variance}.
    \end{itemize}

  \item The analyst may downsample 1-\si{\basepair} resolution signal into bins (see ``Spatial resolution'').
    This involves computing one of:
    \begin{itemize}[nosep,label=\textbullet,align=left,leftmargin=1.5em,labelwidth=1.5em]
    \item Average read count,
    \item Reads per million mapped reads fold enrichment~\cite{larson2013tiered},
    \item Total count of reads~\cite{mammana2015chromatin, taudt2016chromstar, zehnder2019predicting}, or
    \item Maximum count of reads of each bin~\cite{hoffman2012integrative, zhang2016jointly}.
    \end{itemize}
    Binning greatly decreases the computational cost of the \ac{SAGA} algorithm and can improve the data's statistical properties.

  \item The analyst may binarize numeric signal data into presence/absence values~\cite{ernst2012chromhmm, hamada2015learning, biesinger2013discovering, marco2017multi, zhou2016probabilistic}.
    Binarizing signal simplifies analysis by avoiding issues related to the choice of units, but eliminates all but \SI{1}{\bit} of information about signal intensity per bin.
  \end{enumerate}
\end{itemize}

\subsection*{Missing data}

Genomic assays almost always cannot produce signal for every region of the whole genome.
Regions where an assay cannot provide reliable information about the interrogated property constitute ``missing data'' for that assay.
Missing data in sequencing assays may arise due to unmappable sequences, which occur when repetitive reads do not uniquely map to a region~\cite{derrien2012fast, karimzadeh2018}.
Missing data in microarray assays comes from regions covered by no microarray probes.
There are three main ways to treat regions of missing data: (1)~by treating missing data as 0-valued data, (2)~by decreasing the model resolution, averaging over available data so that the missing data has limited impact, or (3)~statistical marginalization over the missing data~\cite{lian2008automated,hoffman2012unsupervised}.

When analyzing coordinated assays across multiple cell types, researchers may have to contend with having no data on some properties within a subset of cell types.
This represents another kind of missing data: one with an entire dataset missing rather than only data at specific positions.
Researchers can impute~\cite{marco2017multi} entire missing datasets through tools such as ChromImpute~\cite{ernst2015large}, PREDICTD~\cite{durham2018predictd} or Avocado~\cite{schreiber2019multi}.
Alternatively they can use \iac{SAGA} model with built-in capability for handling the missing datasets~\cite{biesinger2013discovering}.

\section*{\Acf{HMM} formulation}

Many \ac{SAGA} methods rely on \iac{HMM}, a probabilistic model of the relationships between sequences of observed events and the unobservable hidden states which generate the observed events.
The structure of \acp{HMM}, and similar models such as \acp{DBN}~\cite{dean1989model}, naturally reflect the \ac{SAGA} task of clustering observed data generated by processes that act on sequences of genomic positions.

\subsection*{Simple \acs{HMM} example}
As an illustration of a simple \ac{HMM}, consider a dog, Rover, and his owner, Thomas.
Thomas is 5~years old and too short to see out of the windows in his home.
Rover can leave the house through his dog door and loves taking walks, playing indoors, and napping.
Every morning, he will either wait by the door for Thomas, play with his squeaky toys, or sleep in.
Whichever action he takes depends on the weather he sees outdoors.
For example, on rainy days Rover will more likely nap or play with his toys indoors.

Thomas must infer the state of the weather outside, hidden to him, based on the behavior he observes from Rover.
Thomas knows the general weather patterns outside near his home---for example, that rainy weather likely continues across multiple days.

This scenario fits well into the \iac{HMM} framework.
It has a sequence of observations (Rover's behavior) generated by hidden, non-independent unobservables (the weather outside).
One would like to infer the sequence of hidden unobservables based on the sequence of observations.

\subsection*{Mathematical formulation}

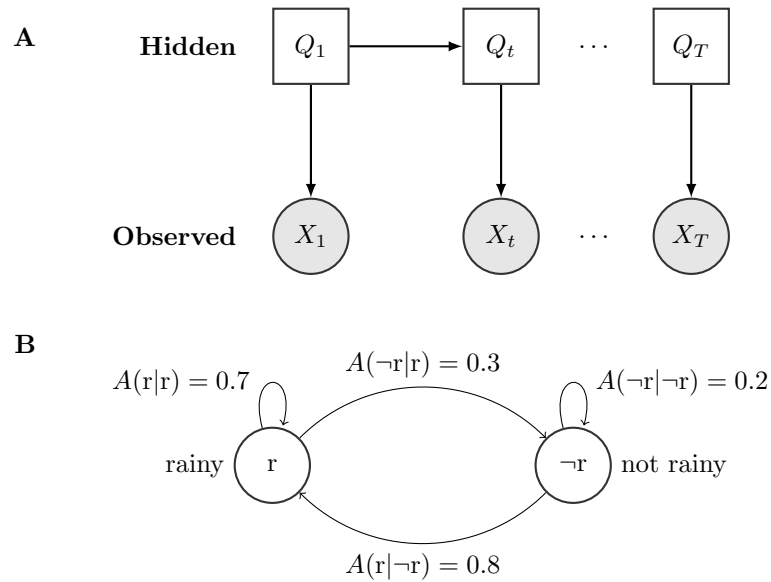
\begin{figure}[htbp]
  \begin{subfigure}[t]{0.08\textwidth}
    \textbf{A}
  \end{subfigure}
  \begin{subfigure}[t]{0.9\textwidth}
    \begin{tikzpicture}[baseline,label distance=1em]
      \node[hidden,label=left:\textbf{Hidden}] (Q1) {$Q_1$};
      \node[hidden] (Qt) [right=of Q1] {$Q_t$};
      \node[hidden] (QT) [right=of Qt] {$Q_T$};

      \node[observation,label=left:\textbf{Observed}] (X1) [below=of Q1] {$X_1$};
      \node[observation] (Xt) [below=of Qt] {$X_t$};
      \node[observation] (XT) [below=of QT] {$X_T$};

      \path (Q1) edge [connect] (Qt)
      (Qt) -- node[auto=false]{\ldots} (QT)

      (Q1) edge [connect] (X1)
      (Qt) edge [connect] (Xt)
      (QT) edge [connect] (XT)

      (Xt) -- node[auto=false]{\ldots} (XT);
    \end{tikzpicture}
  \end{subfigure}

  \vspace{5ex}

  \begin{subfigure}[t]{0.08\textwidth}
    \textbf{B}
  \end{subfigure}
  \begin{subfigure}[t]{0.9\textwidth}
    \begin{tikzpicture}[baseline]
      \node[state,label=left:{rainy}] (Q1) at (1,-1.5) {$\textrm{r}$}
      edge [loop above] node[left] {$A(\textrm{r}|\textrm{r})=0.7$\enspace} ();

      \node[state,label=right:{not rainy}] (Q2) at (5,-1.5) {$\neg{\textrm{r}}$}
      edge [loop above] node[right] {\enspace $A(\neg{\textrm{r}}|\neg{\textrm{r}})=0.2$} ();

      \path (Q1) edge [->,bend left=45] node[above] {$A(\neg{\textrm{r}}|\textrm{r})=0.3$} (Q2)
      (Q2) edge [->,bend left=45] node[below] {$A(\textrm{r}|\neg{\textrm{r}})=0.8$} (Q1);
    \end{tikzpicture}
  \end{subfigure}
  \caption{\textbf{Two representations of \iac{HMM}.
      (a)}~Conditional dependence diagram representation of an unrolled \ac{HMM} with sequence of hidden states~${\{Q_t\}}_{t=1}^T$ and sequence of observations~${\{X_t\}}_{t=1}^T$.
    In this representation, each node represents a hidden discrete~(white rectangle) or observed continuous~(grey circle) random variable.
    For every index~$t$, each hidden random variable~$Q_t$ takes on some value~$q_t$; similarly, each observed variable~$X_t$ takes on some value~$x_t$.
    $X_t$ may represent either scalar or vector observations.
    Solid arcs represent conditional dependence relationships between random variables.
    \textbf{(b)}~State transition diagram representation of Rover and Thomas's weather example.
    In this representation, each node represents a potential value of the hidden variable~$Q_t$.
    The hidden variable takes on values $\textrm{r}$~(rainy) or $\neg{\textrm{r}}$~(not rainy) on any given day~$t$.
    Solid arcs represent transitions between hidden states, which have transition probabilities~$A$.}\label{fig:HMMfig}
\end{figure}

Formally, we can define \iac{HMM} over time~$t \in \{1, \dots, T\}$ as follows~\cite{bilmes2006hmms, yoon2009hidden}.
Let the sequence of observed events~$\boldsymbol{X} = {\{X_t\}}_{t=1}^T$ consist of each observed event~$X_t$ at every time~$t$.
Let the sequence of hidden states~$\boldsymbol{Q} = {\{Q_t\}}_{t=1}^T$ consist of each hidden state~$Q_t$ at every time~$t$.
Each~$Q_t$ takes on a value~$q_t$ from a set of~$m$ possible hidden state values~(\autoref{fig:HMMfig}a).

Under the Markov assumption, the probability of realizing state value~$q_{t+1}$ at the next time step~$t+1$ depends only on the current state value~$q_t$:
\[ P(Q_{t+1} = q_{t+1} | Q_t = q_t, Q_{t-1} = q_{t-1}, \dots, Q_1 = q_1) = P(Q_{t+1}=q_{t+1} | Q_t = q_t). \]
We define the transition probability~$A(q_{t+1} | q_t) = P(Q_{t+1} = q_{t+1} | Q_t = q_t)$, which reflects the frequency of moving from state~$q_t$ to state~$q_{t+1}$.

We define the emission probability~$B(x_t | q_t) = P(X_t = x_t | Q_t = q_t)$ as the probability that the observable~$X_t$ is~$x_t$ if the present hidden state~$Q_t = q_t$.
Specifically, we assume that~$B(x_t | q_t)$ depends only on~$Q_t = q_t$, such that
\[ P(X_t=x_t | Q_t = q_t, Q_{t-1} = q_{t-1}, \dots, Q_1 = q_1) = P(X_t = x_t | Q_t = q_t).\]
Finally, we define the hidden state probability at the first time step as~$\pi_0(q_0) = P(Q_0 = q_0)$.
We can fully define \iac{HMM}~$M = (A, B, \pi_0)$ by specifying all of~$A$, $B$ and~$\pi_0$.

In the case of Rover and Thomas, we have~$m=2$ possible hidden states (rainy, not-rainy) and~$3$ possible observations (Rover is napping, playing indoors, or waiting by the door).
To Thomas, the hidden variable~$Q_t$ captures the weather outside, while the observed variable~$X$ captures Rover's behavior.
The probability of the state of the weather outside changing on a day-to-day basis is defined by the transition probabilities~$A$~(\autoref{fig:HMMfig}b).
The probability of Rover's behavior, given the weather, is defined by the emission probabilities~$B$.

\subsection*{Algorithms for inference, decoding, and training}

\subsubsection*{Inference}
The main task one uses \acp{HMM} for is to quantify how well some predicted sequence of hidden states fits the observed data.
Other common tasks like decoding or training serve as variations of, or build on, this inference task.

In \acp{HMM} inference, we can compute the likelihood of any sequence of hidden states~$\boldsymbol{Q}$.
We use the sequence of observed events~$\boldsymbol{X}$ and the model probabilities~$M$ to compute the likelihood function~$P(\boldsymbol{X}|\boldsymbol{Q}, M)$.
The likelihood function is the probability that our predicted sequence of hidden states produced our observed sequence of observed states.
We often compute the likelihood function using the forward-backward algorithm~\cite{ferguson1980variable,levinson1986continuously}.

\subsubsection*{Viterbi decoding}
Given a sequence of observed events~$\boldsymbol{X}$, we often wish to know the maximum likelihood sequence of corresponding hidden states~$\boldsymbol{Q}$.
For example, if Thomas observes that in the past 3~mornings, Rover slept, played, and then slept again, what weather sequence outside is most likely?

To answer this question, we decode the optimal sequence of hidden states~$\boldsymbol{q}^*$ such that~$\boldsymbol{q}^*=\argmax_{\boldsymbol{Q}} P(\boldsymbol{Q}|\boldsymbol{X},M)$.
The Viterbi algorithm~\cite{viterbi1967error} provides an efficient solution for this problem, making it unnecessary to compare the likelihood for every possible sequence of hidden states.

\subsubsection*{Training}
Usually, we do not know the model parameters~$(A, B, \pi_0)$ and must learn them from data.
We define training as the process of learning these parameters, and training data as the sequence of observations upon which we learn.
An efficient algorithm that finds the global optimum parameter values for some training data does not exist.
Instead, researchers commonly train \acp{HMM} using \ac{EM}~\cite{dempster1977maximum} algorithms such as the Baum-Welch algorithm~\cite{baum1970maximization}, which find a local optimum.
Other reviews~\cite{bilmes2006hmms} describe inference and training methods in more detail.

\subsection*{\acsp{HMM} for \acs{SAGA}}

We can readily apply the \ac{HMM} formalization to genomic data for use in \ac{SAGA} methods.
Instead of time, we define the dynamic axis~$t$ in terms of physical position along a chromosome.
Each position~$t$ refers to a single base pair or, in the case of lower-resolution models, a fixed-size region (see ``Spatial resolution'').
The observation at each genomic position usually represents genomic signal (see ``Input data'').
Each position's hidden state represents its label (see ``Understanding labels'').
As a result, decoding the most probable sequence of hidden states reveals the most probable sequence of labels across the genome.
We call this resulting sequence of labels an annotation.

Many \ac{SAGA} methods use \iac{HMM} structure~\cite{kharchenko2011comprehensive,ernst2012chromhmm,hoffman2012unsupervised,biesinger2013discovering,larson2013tiered,marco2017multi,mammana2015chromatin,poulet2019rt}, or generalizations thereof.
For example, \acp{DBN} are generalizations of \acp{HMM} that can model connections between variables over adjacent time steps.
Methods such as Segway~\cite{hoffman2012unsupervised} use \iac{DBN} model in their approach to the \ac{SAGA} problem.
This can make it easier to extend the model to tasks such as semi-supervised, instead of unsupervised, annotation~\cite{chan2020semi}.

\section*{Understanding labels}

\ac{SAGA} methods are unsupervised.
The labels they produce usually begin with integer designations without any essential meaning.
Ideally each label corresponds to a particular category of genomic element.
To make this correspondence explicit, we must assign a biological interpretation, such as ``Enhancer'' or ``Transcribed gene'', to each label.

Usually, one makes assignments of labels to biological interpretations in a post-processing step.
In post-processing, a researcher compares each label to known biological phenomena and assigns an interpretation that matches the researcher's understanding of molecular biology.
For example, a label characterized by the histone modification H3K36me3 (associated with transcription) and enriched in annotated gene bodies might have the interpretation ``Transcribed''.
A label characterized by H3K27ac and H3K4me1, both histone modifications canonically associated with active enhancers, might have the interpretation ``Enhancer''~\cite{encode2012integrated}.

The interpretation process provides an opportunity to discover new categories of genomic elements.
For example, one \ac{SAGA} study found that their model consistently produces a label corresponding to transcription termination sites.
Previously, none had described a distinctive epigenetic signature for transcription termination~\cite{hoffman2012integrative}.

Manual interpretation proves time-consuming for human analysis.
Applying \ac{SAGA} to multiple cell types independently exacerbates this problem (see ``Annotating multiple cell types'').

Two existing methods automate the label interpretation process: expert rules and machine learning.
In both cases, an interpretation program considers the information that a researcher would use for interpretation.
This includes examining the relationship between labels and individual input data properties.
It also includes reviewing colocalization of labels with features in previously created annotations.
These annotations may have come from \ac{SAGA} approaches or other manual or automated methods.

In the expert rule approach, an analyst designs rules about what properties a given label must have to receive a particular interpretation.
The analyst then applies these rules to assign interpretations to labels from all models~\cite{zerbino2015ensembl}.

In the machine learning approach, one trains a classifier on previous manual annotations.
The classifier then learns a model that assigns interpretations to labels given their properties~\cite{libbrecht2019unified}.
One analysis~\cite{libbrecht2019unified} found that automatic interpretation agreed with manual for 77\% of labels, compared to 19\% expected by chance.

\section*{Spatial resolution}

Baroque music often employs a musical architecture known as ``ternary form''.
Specifically, pieces of this structure follow a general ``ABA'' pattern, whereupon the second ``A'' section recapitulates the first with some variation.
Each section contains multiple musical ``sentences'', which may repeat or vary.
Just like linguistic sentences, each musical sentence contains clusters of notes, or motifs, between ``breaths'' in the musical articulation.
Finer examination of the motifs shows they contain a few notes and chords each.
Finer examination of the notes themselves shows they behave just like isolated phonemes in speech, with little meaning on their own.

The genome resembles a musical composition in that one observes different behaviors at different scales.
The scale of genomic behavior one wishes to observe influences the choice of \ac{SAGA} method and parameters chosen for the method.
To observe nucleosome-scale behavior such as genes, promoters and enhancers, one desires~\SI{\sim 1e3}{\basepair} segments. % chktex 1
To describe behavior on the scale of topological domains~\cite{dixon2012topological}, one desires segments of length \SIrange{1e5}{1e6}{\basepair}~\cite{libbrecht2015entropic,libbrecht2015joint, day2007unsupervised}.

The most important parameter influencing segment length is the underlying resolution of the \ac{SAGA} method.
As noted above (see ``Input data''), most \ac{SAGA} methods downsample data into bins.
To observe nucleosome-scale segment lengths~(\SI{\sim 1e3}{\basepair}), one should use one should use \SIrange{100}{200}{\basepair} resolution~\cite{ernst2012chromhmm, hoffman2012unsupervised, zhang2016jointly}. % chktex 1
To observe domain-scale segment lengths~(\SI{\sim 1e5}{\basepair}), one should use \SI{\sim 1e4}{\basepair} resolution~\cite{filion2010systematic, libbrecht2015joint, wang2020spin}. % chktex 1
Segway~\cite{hoffman2012unsupervised} and RoboCOP~\cite{mitra2020robocop} provide some of few \ac{SAGA} methods optimized for single-base resolution inference, and can identify behavior on a \SI{1}{\basepair} scale.
While most existing \ac{SAGA} methods handle data at just one genomic scale, there exist methods capable of learning from data at multiple genomic scales~\cite{marco2017multi}.

Limitations of of the experimental data itself influence the choice of \ac{SAGA} model resolution.
Spatial imprecision in \ac{ChIP-seq} data gives it an inherent resolution of about \SI{10}{\basepair}.
More precise assays such as ChIP-exo~\cite{rhee2011comprehensive} and ChIP-nexus~\cite{he2015chip} can approach \SI{1}{\basepair} resolution.
Conversely, assays like \ac{DamID} and Repli-seq have a coarser resolution of \SI{\ge 100}{\basepair}. % chktex 1

The desired scale may also influence the choice of input data.
When aiming to annotate at the domain scale, one should include data with activity at this scale, such as replication time data and Hi-C data~\cite{filion2010systematic, libbrecht2015joint, wang2020spin, poulet2019rt}.
The inclusion of long-range contact information from Hi-C data poses a challenge because standard algorithms for \acp{HMM} cannot be used for a probabilistic model that includes long-range dependencies.
Therefore, one must instead use alternative approaches such as graph-based regularization~\cite{libbrecht2015joint} or approximate inference~\cite{wang2020spin}.

\ac{SAGA} methods model segment length through their transition parameters.
\ac{HMM} models assume a geometric distribution in determination of a segment's length~\cite{codogno1987duration}.
Related \ac{DBN} methods can include constraints to tune segment length further.
Constraints include the enforcement of a minimum or maximum segment length~\cite{hoffman2012unsupervised}.
Enforcement of a minimum segment length ensures that one does not obtain segments shorter than the effective resolution of the underlying data or biological phenomena.
Probabilistic models often additionally use a prior distribution on the transition parameters during training to encourage them to produce shorter or longer segment lengths.
% Under this assumption, we can form the transition matrix between hidden states simply by adding the parameters of the Dirichlet prior distribution with the transition matrix implied from the data. These added parameters are known as ``pseudocounts'' for the transition matrix.
% The combination of priors with the matrix implied from the data forms the transition matrix, thereby determining the segment length of the model.
% The summed value of the pseudocounts indicates the strength of the prior and relative distribution of pseudocounts in the matrix indicates the shape of the prior.
% The resulting combined transition matrix defines a geometric distribution of segment length, as before, but the parameters are a mixture of the Dirichlet prior and the transition matrix learned from training data. % Explain; what is relationship to previous part.

% thereby determines the segment length of the model.

\section*{Choosing the number of labels}

% More: mentioned in original outline but discussed: Relationship of number of input datasets to number of labels; ChromHMM original paper's method for choosing number of labels. Max: No space I think.

Most \ac{SAGA} methods require the user to define the number of labels.
Using more labels increases the granularity of the resulting annotation at the cost of added complexity.
Typically, the number of labels ranges from 5--20, with more recent work favoring 10--15 labels.

One might think to make the choice of number of labels automatically with a statistical approach.
The \ac{AIC}, \ac{BIC}, and \ac{FIC}~\cite{fujimaki2012factorized} measure the statistical support a particular number of labels has.
Instead of a fixed number of labels, one may give the model flexibility to choose the number of labels during training and include a hyperparameter that encourages it to choose a higher or lower number~\cite{sohn2015hihmm}.
Or one might define labels according to local minima in an optimization based on a network model of assays~\cite{zhou2016probabilistic}.
One could even exhaustively assign a separate label to every observed presence/absence pattern in binary data~\cite{taudt2016chromstar}.

In practice, however, researchers rarely use these statistical approaches for determining the number of labels.
Optimizing an information criterion does not necessarily yield the most interpretable annotation.
Interpretability reigns supreme in most \ac{SAGA} applications.
End users find annotations most useful when they have about~5--20 labels for two reasons.
First, most can only articulate that many known distinctions between classes of genomic elements.
Second, even if one could find meaningful distinctions between a large number of labels, few using the resulting annotations could keep fine distinctions between such a large number of patterns in their working memory.~\cite{cowan2001magical}
Even if a statistical approach supported the use of 50~labels, the complexity of such an annotation would make it impractical for most users.

\section*{Annotating multiple cell types}

\begin{figure}[htbp]
  \centering
  \includegraphics[width=\textwidth]{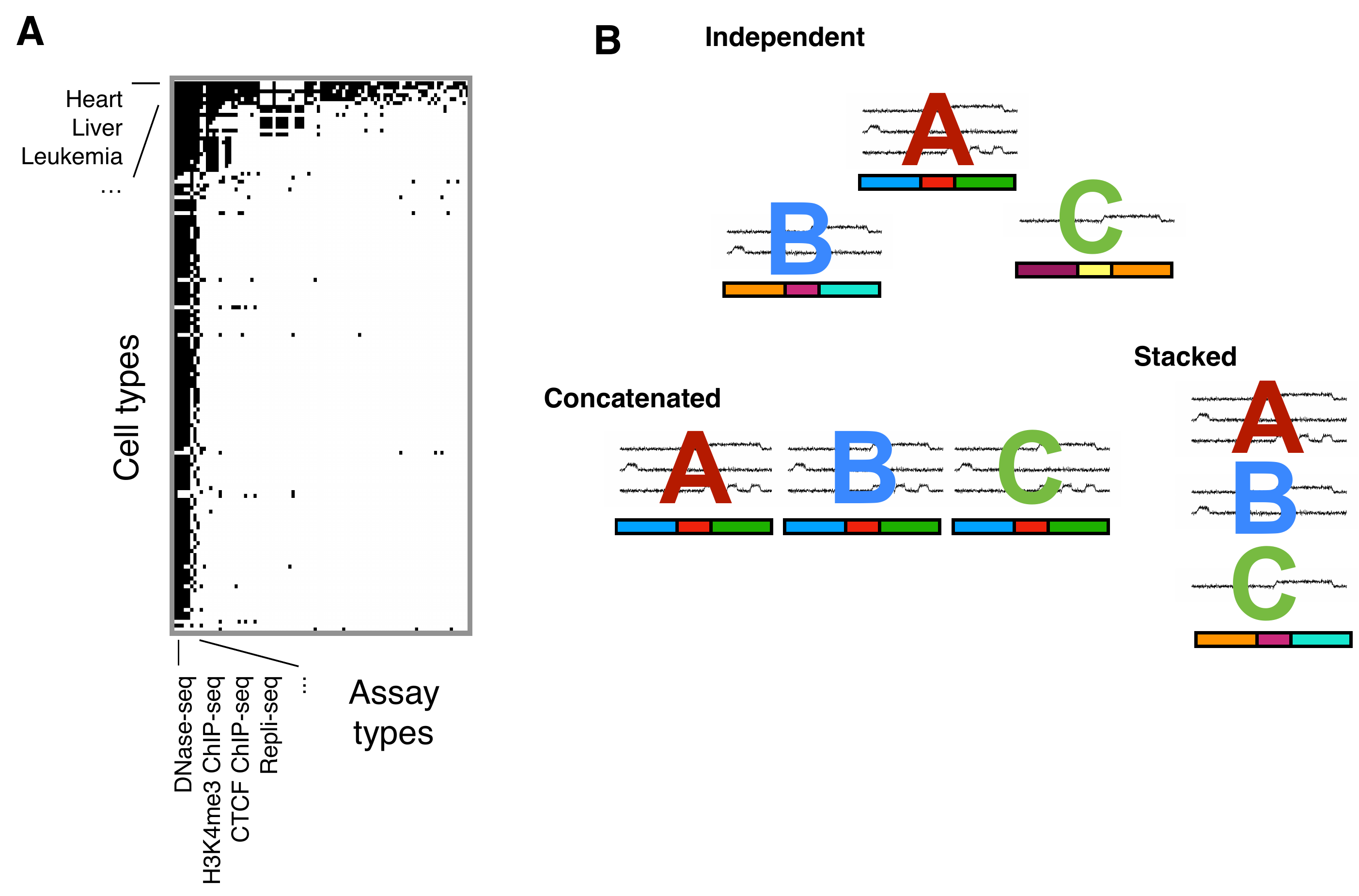} % chktex 8
  \caption{\textbf{Annotating multiple cell types.
    %  (a)}~Datasets generated by the \ac{ENCODE} and Roadmap Epigenomics consortia as of 2019, among NNN cell types and NNN assay types.
    %The NNN black cells represent the NNN datasets actually generated out of NNN combinations of cell type and assay type.
    % Max->Michael. Tracking down the data that produced this plot would be challenging, unfortunately.
      (a)}~Datasets generated by the \ac{ENCODE} and Roadmap Epigenomics consortia as of 2019.
    The black cells represent the datasets actually generated out of a larger number of potential combinations of cell type and assay type.
    \textbf{(b)}~Annotating 6~datasets from 3~different samples: 3~from cell type~A, 2~from cell type B, and 1~from cell type~C.
    Colored letters over signal data indicate data associated with those samples.
    One can use three different \ac{SAGA} strategies with this collection of datasets:
    \emph{Independent}:~Performing training and inference completely independently on each sample.
    This yields a different annotation for each sample.
    \emph{Concatenated} (horizontal sharing):~Training a single model across all cell types. This yields one annotation per sample with a shared label set.
    Each sample must have the same datasets, necessitating imputation of any missing datasets.
    \emph{Stacked} (vertical sharing):~Performing training and inference on datasets from all samples.
    This yields a single pan-cell-type annotation.}\label{fig:multi} % chktex 8
\end{figure}

There now exist epigenomics datasets describing hundreds of biological samples~(\autoref{fig:multi}a).
Researchers have correspondingly adapted \ac{SAGA} methods to work with many samples simultaneously.

We use the term ``sample'' to refer to some population of cells on which one can perform an epigenomic assay.
A sample could correspond to a primary tissue sample, a cell line, cells affected by some perturbation such as drug or disease, or even cells from different species.

The simplest approach for annotating multiple samples involves simply training a separate model on each sample~\cite{libbrecht2019unified}~(\autoref{fig:multi}b).
% Doing so obviates the issues with data availability and artifactual differences discussed below
The large number of models produced by this approach necessitates using an automated label interpretation process (see ``Label interpretation'').

Two categories of approaches aim to share information across samples.
The first, ``horizontal sharing'' or ``concatenated'' approaches, share information between samples to inform the label-training process.
The second, ``vertical sharing'' or ``stacked'' approaches, share position-specific information to inform the label assignment of each position.

% There exist two primary challenges posed by working with multiple samples.
% First, one must perform biological interpretation for every learned model.
% Second, sharing information between samples during training or annotation requires a modification of these methods.
% Two classes of multi-sample \ac{SAGA} methods tackle each of these challenges respectively.

\subsection*{Horizontal sharing: emphasizing similarities across samples for learning labels}

The simplest way to remove the need for interpreting multiple models is to apply a single model across many samples.
To do this, one can treat each sample as referring to separate copies of a longer genome added horizontally after the first one in a ``concatenated'' approach~(\autoref{fig:multi}b).
One performs concatenated training and inference little differently than if the data from different samples pertained to different chromosomes in the same genome.
Because all samples share a single concatenated model, researchers need only perform post-processing interpretation once.

The concatenated approach has wide usage~\cite{ernst2011mapping, hoffman2012integrative, kundaje2015integrative}, but has two downsides.
First, concatenated \ac{SAGA} requires that every sample has data from the same assays.
In practice, this criterion often does not hold true.
This means that---unless these assays are imputed or treated as missing (see ``Vertical sharing: emphasizing similarities across samples in positional information'')---one must exclude data for an assay conducted in even all but one samples.
In a simple concatenated approach, one cannot annotate a sample which lacks even one dataset present in the others.

Second, data from different samples can have artifactual differences or batch effects.
Applying the same model across multiple cell types assumes that all datasets from the same assay type have similar statistical properties.
This can result in label distributions to vary wildly across samples and biologically implausible sample-specific labels.
Data normalization can help abate the problem of different statistical properties between samples, but usually does so incompletely.
This problem is particularly significant when using continuous signal.
In contrast, binarizing the data (see ``Input data'') can cover up some experimental biases.

% Max: cite the ENCODE cross-species chromatin (http://doi.org/10.1038/nature13415, ho2014comparative) as an interesting example of concatenated. Max->Michael: Any particular point you want to make about it?

One might expect that concatenated annotation would benefit training by increasing the amount of training data.
As it turns out, multiplying the amount of training data does not significantly aid the training process, as the types of labels vary little across samples.
Most complex eukaryotic organisms studied with \ac{SAGA} have very large genomes, and just one sample provides plenty of training data.
In fact, for computational efficiency, researchers often train on just a fraction of the available samples~\cite{kundaje2015integrative}, a fraction of the genome from a given sample~\cite{hoffman2012unsupervised} or both.

% Max: should discuss that the concatenated approach throws out position-specific information

% Max: Discussion of the post-processing automated approach in the Segway encyclopedia more?

\subsection*{Vertical sharing: emphasizing similarities across samples in positional information}

Another class of multi-sample \ac{SAGA} methods shares position-specific information across samples as part of the annotation process.
These methods take advantage of the non-independence of biological activity across samples at a genomic position.
For example, if a given position has an active enhancer label in many samples, it is more likely to act as an active enhancer in a new sample.

The simplest type of vertical sharing approach learns a model on data from all samples jointly~(\autoref{fig:multi}b).
One can implement this ``stacked'' approach by adding datasets from all samples vertically into a single combined model.
A stacked model captures patterns of activity across multiple cell types.
For example, a stacked model, unlike an independent model, can find a label for an enhancer active in cell type~A and cell type~B but inactive in cell type~C.

Although conceptually simple, the stacked approach tends not to work well for more than several cell types.
Stacking fails with larger number of cell types because each pattern of activity requires its own label.
Therefore, the number of labels must grow exponentially in the number of samples.
A simple stacked model that treats all assays as independent also ignores the relationship between assays on the same cell type, or the same assay type on different cell types.

A second approach uses a concatenated model that additionally learns a position-specific preference over the labels for each position.
Through this preference, data from one sample can influence inference on another.
Two implementations have applied variants of this hybrid horizontal-vertical sharing approach.
First, TreeHMM~\cite{biesinger2013discovering} uses a cellular lineage tree as part of its input.
For each genomic position, TreeHMM models statistical dependency between the label of a parent cell type and that of a child cell type.
Second, IDEAS~\cite{zhang2016jointly} uses a similar approach to TreeHMM, but dynamically identifies groups of related samples rather than using a fixed developmental structure.
The IDEAS model allows these sample groups to vary across positions, which allows for different relationships between samples in different genomic regions.

A third approach for vertical sharing uses a pairwise prior to transfer position-specific information between cell types~\cite{libbrecht2015entropic,libbrecht2015joint}.
In other words, if position~$i$ and position~$j$ received the same label in many other samples, then they should be more likely to receive the same label in an additional sample.
In contrast to the other methods in this section, the pairwise prior approach does not require the use of concatenated annotation.
Therefore, the pairwise approach has the advantage of not requiring the same available data in all cell types.

A fourth approach imputes missing datasets in the target cell type, then applies any of the above annotation methods to the imputed data~\cite{ernst2015large}.
Imputation provides three advantages.
First, it ensures that all target cell types have the same set of datasets.
Second, one can conduct imputation entirely as a preprocessing step, allowing the use of any \ac{SAGA} method afterward.
Third, the imputation process can normalize some artifactual differences between datasets, making concatenated annotation more appropriate.

Vertical sharing approaches have the downside that one cannot understand the annotation of each sample in isolation.
This arises from the influence on label assignments in one sample by data from other samples.
In particular, vertical sharing tends to mask differences between samples.
For example, if some position has an enhancer label in many samples, vertical sharing approaches will annotate that position as an enhancer in a target cell type too, even with no enhancer-related data in the target cell type.

\section*{Using and visualizing \acs{SAGA} annotations}

\begin{table}[htbp]
  \begin{adjustwidth}{-2.25in}{0in}
  \begin{tabularx}{\linewidth}{lXrrrr}
    \hline % chktex 44
    \textbf{Name} & \textbf{Method} & \textbf{Labels} & \textbf{Samples} & \textbf{Datasets per sample} & \textbf{Reference} \\
    \hline % chktex 44
    Roadmap (5 mark) & ChromHMM concatenated & 15 & 127 & 5 & \cite{kundaje2015integrative} \\ % chktex 2
    Roadmap (6 mark) & ChromHMM concatenated & 18 & 98 & 6 & \cite{kundaje2015integrative} \\ % chktex 2
    Roadmap (imputed) & Impute+ChromHMM & 25 & 127 & 5--12 & \cite{kundaje2015integrative} \\ % chktex 2
    IDEAS & IDEAS & 20 & 127 & 5 & \cite{zhang2017accurate} \\ % chktex 2
    Segway Encyclopedia & Segway independent & 14--33 & 164 & 3--126 & \cite{libbrecht2019unified} \\ % chktex 2
    Segway domains & Segway+GBR & 5 & 8 & 12 & \cite{libbrecht2015joint} \\ % chktex 2
    Ensembl Regulatory Build & ChromHMM independent & 25 & 89 & 6--10 & \cite{zerbino2015ensembl} \\ % chktex 2
    \hline % chktex 44
  \end{tabularx}
  \caption{\textbf{Existing large-scale human reference annotations.}}\label{tab:reference}
  \end{adjustwidth}
\end{table}

A number of resources can aid in the application of \ac{SAGA} algorithms and annotations.
Reference annotations exist for many cell types~(\autoref{tab:reference}).
These obviate the need for a user of the annotation to actually run \iac{SAGA} method.
Alternatively, if the user must run \iac{SAGA} algorithm on their own data, standardized protocols describe how to perform this process~\cite{roberts2016semi, ernst2017chromatin}.

\begin{figure}[htbp]
  \centering
  \includegraphics[width=\textwidth]{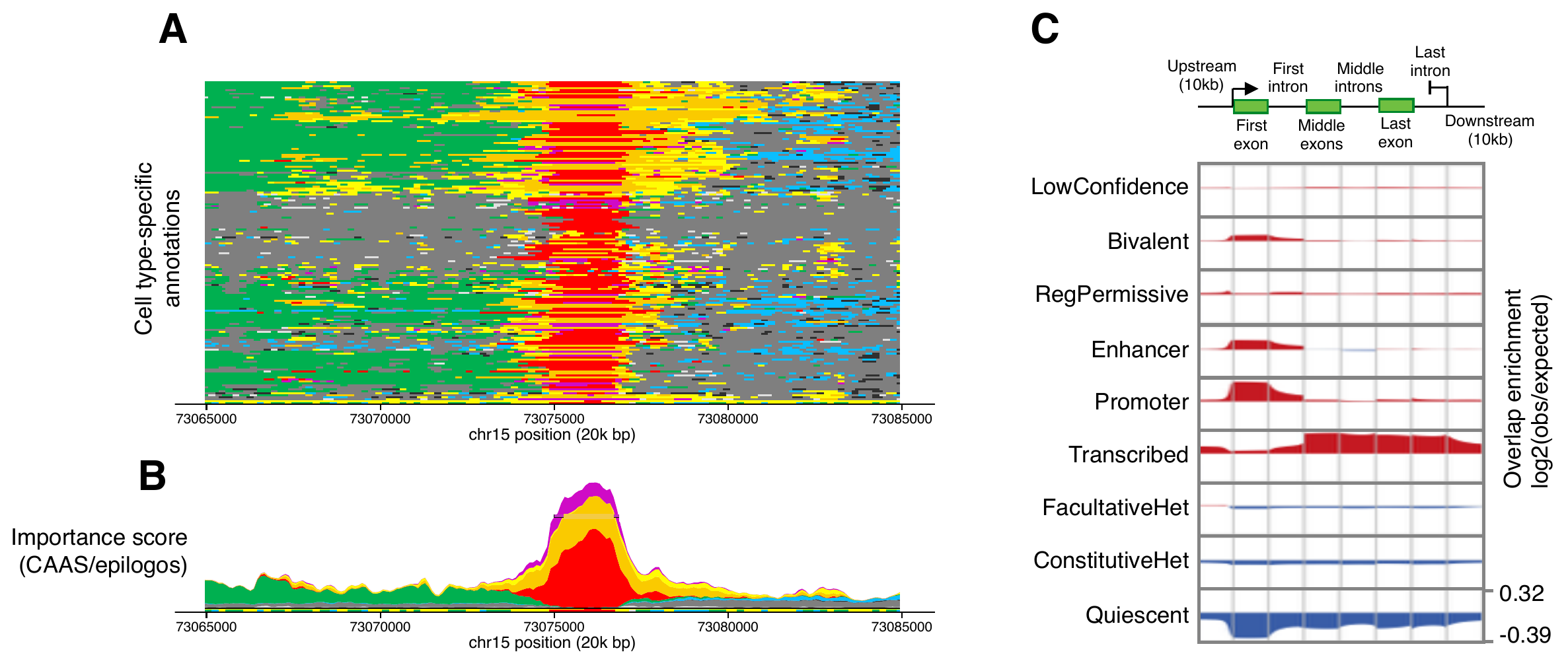}
  \caption{
  \textbf{Visualizations of \ac{SAGA} annotations.
      (a)}~Genome browser display showing 164~cell type annotations for a \SI{20}{\kilo\basepair} region on human chromosome~15 (GRCh37/hg19)~\cite{church2011modernizing}.
    Each annotation has 8~labels: Promoter~(red), Enhancer~(orange), Transcribed~(green), Permissive regulatory~(yellow), Bivalent~(purple), Facultative heterochromatin~(light blue), Constitutive heterochromatin~(black), Quiescent~(grey), and Low Confidence~(light grey).
    \textbf{(b)}~Importance score (\ac{CAAS}) for the same region.
    Total height at each position indicates the position's estimated importance.
    Height of a given color band denotes the contribution towards importance of the associated label.
    \textbf{(c)}~Genome-wide visualization of the \ac{SAGA} annotation for 164 samples aggregated over GENCODE~\cite{frankish2019gencode} protein-coding gene components.
    Rows:~the 9 labels of the annotation.
    Columns:~gene components, including \SI{10}{\kilo\basepair}~flanking regions upstream and downstream.
    Each cell shows the enrichment of the row's label with a position along the column's component.
    Figures derived from~\cite{libbrecht2019unified}.}\label{fig:visualization}
%   Max: If you displayed a different region for subfigs A and B, it would demonstrate that the CAAS is useful at more than just one region.
\end{figure}

Most users of \ac{SAGA} annotations view them through one of three visualization strategies.
The first, and most common, strategy displays individual annotations as individual rows or ``tracks'' on a genome browser~(\autoref{fig:visualization}a).
In each row, the browser displays the segments of that annotation for a region of one chromosome, usually indicating the label by color.
Popular genome browsers for displaying segmentations include the \ac{UCSC} Genome Browser~\cite{kent2002human}, the \ac{WashU} Epigenome Browser~\cite{zhou2011human}, and Ensembl~\cite{zerbino2017ensembl}.

A second visualization strategy integrates annotations of all samples~(\autoref{fig:visualization}b).
This visualization stacks all labels for a given position on top of one another and scales the vertical axis by an estimate of functional importance of that position.
Two methods can estimate this importance: Epilogos~(\url{https://epilogos.altius.org/}), which emphasizes rare activity, and the \ac{CAAS}, which measures activity that is correlated with evolutionary conservation~\cite{libbrecht2019unified}.

A third visualization strategy aggregates information about each label across the entire genome.
This shows the enrichment of each label at positions of known significance, such as gene components~(\autoref{fig:visualization}c) or curated enhancers.
Tools such as Segtools~\cite{buske2011exploratory} and deepTools~\cite{ramirez2014deeptools} can create these visualizations.

\ac{SAGA} annotations can provide valuable reference datasets to other analyses and tools.
The assignment of one and only one label from a small set to every mappable part of the genome greatly eases downstream analyses.
\ac{SAGA} annotations summarize genomic activity in a much simpler way than the individual input datasets, and even than processed versions of the input datasets such as peak calls.

Most \ac{SAGA} annotations are in the tab-delimited \ac{BED} format~(\url{https://genome.ucsc.edu/FAQ/FAQformat.html#format1}).
This makes it easy to remix \ac{SAGA} annotations with other datasets using powerful software such as BEDTools~\cite{quinlan2010bedtools}.
\ac{SAGA} annotations form building blocks for methods for integrative analysis of genomic data such as CADD~\cite{kircher2014general}.

\section*{Conclusions and outlook for future work}

\ac{SAGA} methods provide a powerful and flexible tool for analyzing genomic datasets.
These methods promise to continue to play an important role as researchers generate more datasets.
Despite the large existing literature, future work could still address many challenges.

\paragraph{Alternate scales and data types.}
Nucleosome-scale annotations (\SIrange{100}{1000}{\basepair} segments) of histone modifications have wide usage.
While annotations of different data types or at different length scales exist, they are less widely used.
Currently, there exist reference domain annotations with segments of length \SIrange{1e5}{1e6}{\basepair} for only a small number of samples~\cite{filion2010systematic, larson2013tiered, rao20143d, libbrecht2015joint}, and few or no annotations at other scales.

\paragraph{Data preprocessing}.
Genome annotations would improve with better processing and normalization of input datasets.
Representations such as fold enrichment used by existing methods seem primitive compared to more rigorous quantification schemes used in RNA-seq analysis such as \ac{TPM}.
One could also improve \ac{SAGA} preprocessing by more frequently incorporating information from multi-mapping reads~\cite{zeng2015perm}.

\paragraph{Confidence estimates.}
Most methods do not report any measure of confidence in their predictions.
Two types of confidence would prove useful.
First, one would often like to know the level of confidence that a position in some sample has label~X and not label~Y.
Second, in many cases one would like to have confidence in a differential labeling between two samples---that cell type~A and cell type~B have different labels.
Two methods work towards a solution for the second problem~\cite{yen2015systematic, ebert2018fast}, but there remains much room for further work.

\paragraph{Continuous representations.}
Existing \ac{SAGA} methods output a discrete annotation, assigning a single label to each position.
In this discrete approach, annotations cannot easily represent varying strength in activity of genomic elements or elements that simultaneously multiple types of activity.
A continuous annotation approach analogous to the topic models used for text document classification might address this limitation~\cite{chen2018continuous}.

\paragraph{Single cell data.}
Existing \ac{SAGA} methods use data from bulk samples of cells.
Increasing availability of data from single-cell assays necessitates the development of methods that can leverage this additional information.

\paragraph{Pan-cell-type annotation.} % chktex 8
The semantics of genome annotations correspond poorly to the way most molecular biologists conceptualize genomic elements.
Most existing annotations are cell-type-specific---the annotation states that a given locus acts as an active enhancer in cell type~A. % chktex 8
In contrast, researchers often state that a given locus ``is an enhancer''.

In contrast, other annotations---such of those of protein-coding genes---serve as a pan-cell-type characterization. % chktex 8
Each gene has a fixed location, and only its expression varies across samples.

There exists a need for pan-cell-type epigenome annotations. % chktex 8
Such an annotation would define fixed intervals for regulatory elements such as promoters, enhancers, and insulators, and it would specify in which samples each element is active.
Specifically targeting this task in the \ac{SAGA} model could improve results over pan-cell-type annotations assembled from multiple cell-type-specific \ac{SAGA} models~\cite{libbrecht2019unified}. % chktex 8

\section*{Author contributions}

Conceptualization, M.W.L. and M.M.H.;
Funding acquisition, M.M.H.;
Investigation, M.W.L., R.C.W.C., and M.M.H.;
Visualization, M.W.L. and R.C.W.C.;
Writing --- original draft, M.W.L., R.C.W.C., and M.M.H.;
Writing --- review \& editing, M.W.L. and M.M.H.

\section*{Acknowledgments}

This work was supported by the Natural Sciences and Engineering Research Council of Canada (RGPIN-2015-03948 to M.M.H.) % chktex 8

\section*{Competing interests}

The authors declare no competing interests.

%\nolinenumbers{}

% Either type in your references using
% \begin{thebibliography}{}
%   Text
% \end{thebibliography}
%
% or
%
% Compile your BiBTeX database using our plos2015.bst
% style file and paste the contents of your .bbl file
% here. See http://journals.plos.org/plosone/s/latex for
% \begin{thebibliography}{10}

% \bibitem{bib1}
%   Conant GC, Wolfe KH.
%   \newblock {{T}urning a hobby into a job: how duplicated genes find new
%   functions}.
%   \newblock Nat Rev Genet. 2008 Dec;9(12):938--950.

% \bibitem{bib2}
%   Ohno S.
%   \newblock Evolution by gene duplication.
%   \newblock London: George Alien \& Unwin Ltd. Berlin, Heidelberg and New York:
%   Springer-Verlag.; 1970.

% \bibitem{bib3}
%   Magwire MM, Bayer F, Webster CL, Cao C, Jiggins FM.
%   \newblock {{S}uccessive increases in the resistance of {D}rosophila to viral
%   infection through a transposon insertion followed by a {D}uplication}.
%   \newblock PLoS Genet. 2011 Oct;7(10):e1002337.

% \end{thebibliography}

\section*{Acronyms}

\begin{acronym}
  \acro{AIC}{Akaike information criterion}
  \acroindefinite{AIC}{an}{an}

  \acro{ATAC-seq}{assay for transposase-accessible chromatin-sequencing}
  \acroindefinite{ATAC-seq}{an}{an}

  \acro{BED}{browser extensible data}
  \acro{BIC}{Bayes information criterion}
  \acro{CAAS}{conservation-associated activity score}
  \acro{ChIP-seq}{chromatin immunoprecipitation-sequencing}
  \acro{CUTnRUN}[CUT\&RUN]{cleavage under targets and release using nuclease}
  \acro{DamID}{DNA adenine methyltransferase identification}

  \acro{DNase-seq}{deoxyribonuclease-sequencing}
  \acro{DBN}{dynamic Bayesian network}

  \acro{EM}{expectation-maximization}
  \acroindefinite{EM}{an}{an}

  \acro{ENCODE}{Encyclopedia of DNA Elements}
  \acroindefinite{ENCODE}{an}{an}

  \acro{FIC}{factorized information criterion}
  \acroindefinite{FIC}{an}{a}

  \acro{HMM}{hidden Markov model}
  \acroindefinite{HMM}{an}{a}

  \acro{SAGA}{segmentation and genome annotation}
  \acro{TPM}{transcripts per million}

  \acro{UCSC}{University of California, Santa Cruz}

  \acro{WashU}{Washington University in St.~Louis}
\end{acronym}

%\bibliography{bibliography}

\end{document}